# Heat-tree: Cross-platform software for interactive and embeddable phylogenetic tree visualization and editing

Zachary S.L. Foster[1], Jeff H. Chang[2], and Niklaus J. Grünwald[1]

[1] Horticultural Crops Disease and Pest Management Research Unit, USDA Agricultural Research Service, Corvallis, OR 97330, USA
[2] Department of Botany and Plant Pathology, Oregon State University, Corvallis, OR 97331, USA

**Corresponding author:** Zachary S.L. Foster (zachary.foster@usda.gov)

## Abstract

Phylogenetic trees are the primary framework for conveying evolutionary relationships. While many tools exist for visualizing phylogenetic trees, most are limited to static graphics, require coding expertise, or are developed for a specific website and not easily reusable or extensible. To address these limitations, we developed `heat-tree`, a collection of software packages in JavaScript, R, and Python for interactive visualization, manipulation, and editing of phylogenetic trees and their associated metadata. `Heat-tree` allows for the creation of customizable, web-compatible tree visualizations that can be easily embedded in R Markdown, Jupyter Notebooks, and Quarto documents, as well as directly in HTML/JavaScript applications and websites. The package supports radial and rectangular tree layouts, automated translation of metadata values into visual encodings on the tree, interactive tree editing, and export capabilities for publication-quality figures. All visualization parameters are definable programmatically or interactively using the comprehensive graphical user interface included with each visualization. `Heat-tree` was designed to be a user-friendly software package for interactive tree viewing, manipulation, editing, and self-contained, embeddable visualization across software environments.

## Interpretive Summary

Phylogenetic trees are widely used to study evolutionary relationships among organisms, yet existing computational tools for viewing and working with these trees are often difficult to use or limited in flexibility. We developed `heat-tree` to address these challenges. This software provides an integrated user-friendly method for interactive tree visualization and editing. Researchers can link biological data directly to trees and produce custom, web-compatible visualizations. By making phylogenetic tree visualizations more flexible, reusable, and accessible, `heat-tree` lowers technical barriers and enables broader use of evolutionary analyses in research, education, and communication.

## Main text

Phylogenetic trees are the primary tool used to plot biological relationships. Beyond representing evolutionary relationships, trees can be used as hierarchical backbones to visualize diverse biological data, including traits, abundance measurements, and experimental treatments, by encoding metadata as color, size, or other visual attributes. These tree-based visualizations require coordinated manipulation of both the tree itself and associated metadata. Tree data are commonly represented as directed acyclic graphs, while associated metadata are stored in tabular formats with a column mapping rows to tree node identifiers. Existing software programs designed to manipulate and visualize this combination of data structures either require coding expertise, rely on Graphical User Interfaces (GUIs) that cannot be controlled programmatically, or cannot be readily reused or embedded in other contexts.

Existing phylogenetic visualization software can be broadly categorized into two groups: software with GUIs such as FigTree (Rambaut 2009), iTOL (Letunic and Bork 2021), and Dendroscope (Huson and Scornavacca 2012), and programming library-based approaches including the R packages ape (Paradis and Schliep 2019), ggtree (Yu et al. 2017), and phytools (Revell 2012) (Table 1). These two categories of software each have distinct strengths and limitations, which impose a trade-off between ease of use and flexibility. GUI-based applications typically offer user-friendly interfaces but cannot be embedded into other tools or reports. Conversely, programming-based solutions provide flexibility, reproducibility, and embeddability but require substantial coding expertise and typically produce visualizations that lack interactivity. There is a need for tools that combine ease of use with the flexibility required for reproducible and customizable phylogenetic visualization in various programming environments. We previously developed the metacoder R package, introducing the concept of "heat trees," in which statistical data are visualized by encoding them onto taxonomic trees using node sizes and colors (Foster et al. 2017). Here we extend this visualization concept to interactive phylogenetic trees through a coordinated suite of tools, including a `heat-tree` JavaScript library, the `heattree` R package, and the `heattree_py` Python package. For simplicity, we will refer to `heat-tree,` the name of the JavaScript library, when the discussing the collection of software in general.

`Heat-tree` is a multi-platform, interactive visualization tool for creating phylogenetic trees embedded in both web-based and computational workflows. Website and application developers can use the JavaScript library directly, while R and Python users can seamlessly integrate the interactive web-compatible visualization with phylogenetic tools available in those languages. This allows figures to be generated as part of reproducible analyses without requiring specialized software or web design experience. For example, rendering R Markdown or Quarto documents to HTML enables reports and websites with embedded interactive trees while the Python package supports similar workflows within interactive Jupyter Notebooks. To support this design, the core visualization engine is implemented in JavaScript as the `heat-tree` library, available from NPM (https://libraries.io/npm/heat-tree). This library provides all functionalities required for web-based applications, including rendering, layout algorithms, and interactive manipulation and editing. The `heattree` R package adapts the JavaScript library for convenient use within the R environment (currently under review at CRAN). The `heattree_py` package provides equivalent functionality for Python-based workflows and is available on PyPI (https://pypi.org/project/heattree-py).

The comprehensive GUI built into `heat-tree` provides point-and-click access to all supported features. A collapsable hierarchical tool bar provides access to global controls and contextual buttons appear when tree features are selected. Trees and associated metadata can be loaded interactively from files and users can switch between multiple trees, each with its own metadata mappings and settings. Users can customize automated zooming and panning behavior, modify tree layouts and branch length, and adjust the mapping between metadata columns and visual attributes (e.g. label color and node size), and customize color palettes. Interactive selection of clades allows for them to be removed or rotated and for subtrees to be collapsed or expanded to replace the current tree. Together, these controls allow users to fine-tune visualization directly within the rendered tree and export figures at the desired format and resolution without having to modify source code. Any instance of the visualizations produced by `heat-tree`, including those provided as examples in the online documentation, can be used to import new trees and metadata, customize their appearance, and generate publication-quality figures through point-and-click interaction alone.

The JavaScript `heat-tree` library accepts phylogenetic trees in Newick or Nexus format and associated metadata in Tab/Comma-Separated Value (TSV/CSV) format. Because JavaScript running in a web browser does not generally have direct access to the filesystem, tree and metadata inputs are provided as text at the time of initiation. In practice, data retrieval is typically handled by surrounding application code, such as loading data from user input, a file upload, or a database. However, once a visualization is initiated, users can interactively upload new tree and metadata files through the embedded GUI. Each visualization is rendered within a specified Document Object Model (DOM) element, which determines where the interactive tree appears in the HTML document. In the following code, “#container” designates where the tree visualization is placed and demonstrates how a minimal HTML file can be used to generate a heat-tree visualization from a small example dataset.

```
<!DOCTYPE html>
<div id='container' style='width:800px;height:300px;'></div>
<script type='module'>
  import {heatTree} from 'https://esm.sh/@grunwaldlab/heat-tree';
  const newick = '(A:0.1,B:0.2,(C:0.3,D:0.4):0.5);';
  const metadata = `
node_id\tabundance\tsource
A\t145\tfarm
B\t892\tnursery
C\t234\tcity
  `;
  heatTree(
    '#container',
    {
      tree: newick,
      metadata: metadata,
      aesthetics: {
        tipLabelColor: 'source',
        tipLabelSize: 'abundance'
      }
```

```
    }
  );
</script>
```

This JavaScript code shows a small illustrative dataset, composed of a simple Newick-formatted phylogenetic tree and associated metadata, which are passed to `heat-tree` to generate an annotated interactive visualization (Figure 1). In this example, the “source” column is used to determine the color of tip labels, and the “abundance” column is used to determine the size. Note how many aspects of the plot do not need to be defined manually in this code, such as the scaling of branches, font sizes, color palettes, whether a column contains categorical or continuous data, or which column in the metadata contains the node IDs. All these and more features are automatically determined when not specified.

The R and Python packages build upon the JavaScript library by providing language-native interfaces for specifying data, settings, and metadata mappings. Each package accepts inputs in formats natural to its respective language, translates inputs to the corresponding JavaScript command, and generates self-contained HTML that bundles the JavaScript library with the respective commands. The resulting interactive widget is then displayed automatically in a manner appropriate to the execution context, such as an interactive R/Python/Jupyter Notebook session or a rendered Quarto, R Markdown, or Jupyter Notebook document. This allows visualizations embedded in such documents to work offline. During interactive use, both packages will automatically display trees in a browser or the plot viewer pane of interactive development environments (IDEs) like RStudio. Together, these features allow R and Python users to leverage interactive JavaScript-based graphics with more familiar workflows. Both the R and Python packages include example data from Weisberg *et al.* (2020) to allow users to easily test the packages and these data are used in the examples below (Figure 2).

The R package `heattree` accepts tree data as either file paths or as `phylo` objects from the widely used `ape` package, along with metadata provided as file paths or as `data.frames/tibbles`. This R example uses the contents of the “strain” column for tip labels and the contents of the “host_type” column to color tip labels:

```
library(heattree)
heat_tree(
  tree = weisberg_2020_mlsa,
  metadata = weisberg_2020_metadata,
  aesthetics = c(
    'tipLabelText' = 'strain',
    'tipLabelColor' = 'host_type'
  )
)
```

For the Python package, tree data can be provided as file paths, raw strings, `ete3 TreeNode` objects, `dendropy Tree` objects, or Biopython `Phylo` objects. Metadata are accepted as file paths or `pandas DataFrame` objects:

```
from heattree_py import heat_tree, example_data
```

```
heat_tree(
    tree = example_data('weisberg_2020_mlsa'),
    metadata = example_data('weisberg_2020_metadata'),
    aesthetics = {
      'tipLabelText': 'strain',
      'tipLabelColor': 'host_type'
    }
)
```

An important feature of `heat-tree` is the flexibility in mapping metadata columns to visual aesthetics. Mapping uses a consistent syntax analogous to that of ggplot2 (Wickham 2016), where column names in user-specific metadata files are assigned to different "aesthetics" (e.g. tip-label colors). The software then infers appropriate transformations of the associated data to aesthetic values, like color and component sizes. Metadata tables include a column of tip identifiers that match those in the Newick tree and the software automatically identifies this column. Both continuous and categorical metadata are supported, with mappings summarized in an automatically generated legend that shows how values correspond to aesthetics. Color palettes associated with these aesthetics can be adjusted interactively, allowing users to refine the appearance of the visualization as needed.

Another useful feature of `heat-tree` is that it automatically scales and lays out trees to adapt the available plotting space and resolution. As a result, a single heat-tree specification produces appropriate visualizations across different display contexts without manual reconfiguration. This is accomplished using a combination of the D3 JavaScript package and space-optimization algorithms. Tree layout and scaling are computed from a mathematical representation of the tree that jointly determines branch-length scaling and label sizing based on the available plotting space. These parameters are optimized using a series of constraints that balance readability, proportional branch lengths, and efficient use of the viewing window. Constraint priorities reflect common visualization requirements, including readable text at 100% zoom, sufficient branch length relative to label overhang, overall fit within the viewing window, visible lengths for the shortest branches, and an upper bound on text size. Once suitable scaling values are determined, all other visual properties of the tree can be inferred. Importantly, this model allows the final size of the visualization to be known in advance, enabling smooth automatic zooming, panning, and animated transitions.

In summary, `heat-tree` was developed to provide flexibility, reproducibility, and accessibility for interactive phylogenetic tree visualization. It supports the production of publication-quality, interactive trees that can be embedded directly in documents generated from data analysis workflows. Its GUI and multi-platform support lowers barriers to adoption and enables use of phylogenetic tree visualizations by researchers across disciplines with varying levels of technical expertise. The package's extensive customization options, metadata integration capabilities, and seamless bioinformatic workflow integrations make it valuable for phylogenetic analysis across diverse biological disciplines. We recently incorporated `heat-tree` into the Nextflow pipeline *PathogenSurveillance*, which automates genome analysis and produces phylogenetic trees as outputs (Foster et al. 2025). By integrating `heat-tree`, these trees can be interactively visualized, analyzed, and refined in output reports, rather than treated as static end products, enabling rapid iteration and more direct generation of publication-quality figures. As biological datasets continue to grow in complexity, tools such as `heattree` that facilitate analysis and communication of phylogenetic patterns will become increasingly valuable.

**Acknowledgments**

We thank the members of the Chang, LeBoldus, Weisberg, and Grunwald labs for helpful insights and feedback. We also thank the JavaScript, R, and Python communities and contributors to the `D3` and `htmlwidgets` packages, which provide essential functions for `heattree`. All intellectual content, including conceptualization, methodology, data generation, and decision making, were the responsibilities of the authors. An AI tool was used only for grammar, spelling, and general text polishing of the manuscript. Large language models (kimi-k2.5 and claude-sonnet-4-5) were used via Aider and OpenCode during the development of `heat-tree`, `heattree`, and `heattree_py`.

**Data and code availability:**

The JavaScript library `heat-tree` is available on npm at https://libraries.io/npm/heat-tree, the R package was submitted to CRAN, and the Python package is available on PyPI at https://pypi.org/project/heattree-py/. Information on installation, tutorials, and API references can be found on each package's respective documentation website: https://grunwaldlab.github.io/heat-tree/ for the JavaScript library, https://grunwaldlab.github.io/heattree for the R package, and https://grunwaldlab.github.io/heattree_py for the Python package. The source code for each package can be found on GitHub: https://github.com/grunwaldlab/heattree for the JavaScript library, https://github.com/grunwaldlab/heattree for the R package, and https://github.com/grunwaldlab/heattree_py for the Python package. A prebuilt binary for the R package is also distributed through Bioconda as `r-heattree`, facilitating reproducible installations within conda environments. `Heat-tree` is released under the MIT License. Community feedback and contributions are welcomed via GitHub.

**Funding:**

This work was supported by projects and grants from the USDA Agricultural Research Service (2072-22000-045-000-D) and USDA NIFA (2021-67021-34433).

## Figures

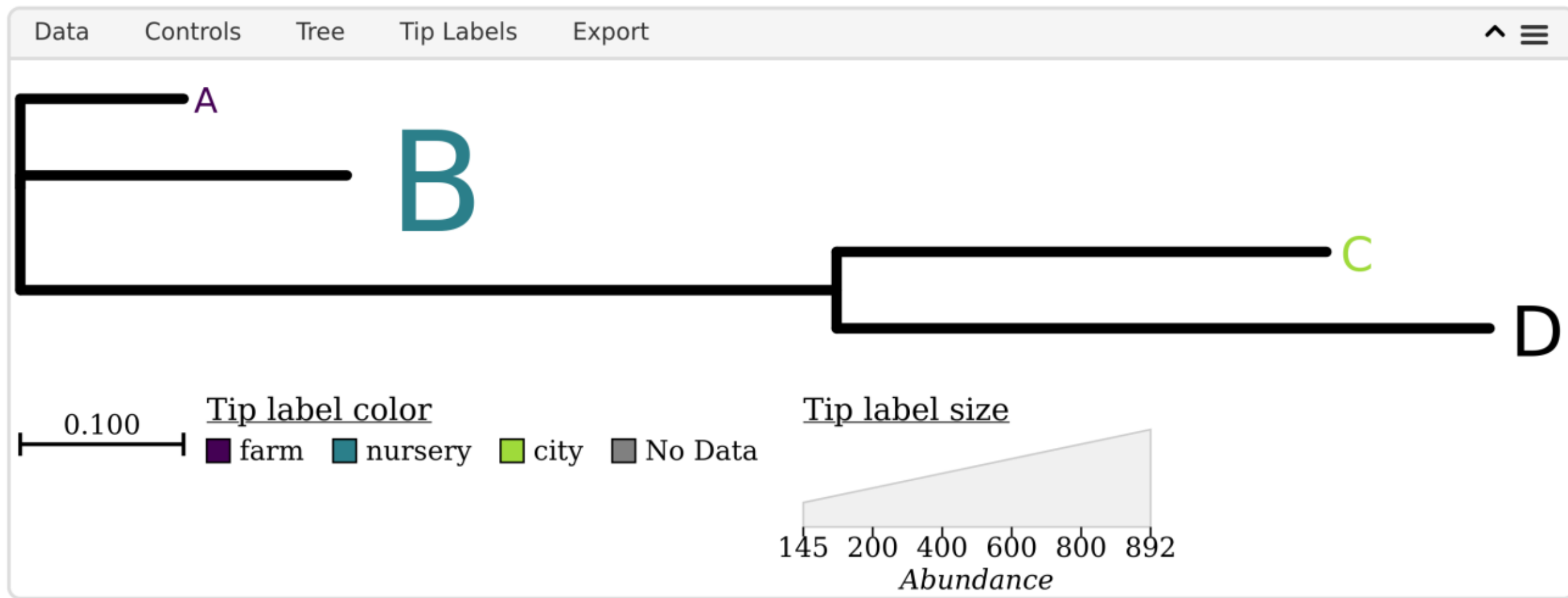

**Figure 1. Interactive phylogenetic tree visualization produced with** `heat-tree` **implemented in JavaScript.** Here, tip label color is determined by a column with categorical data and tip label size is determined by a column with continuous numerical data. Colors and sizes used are automatically determined.

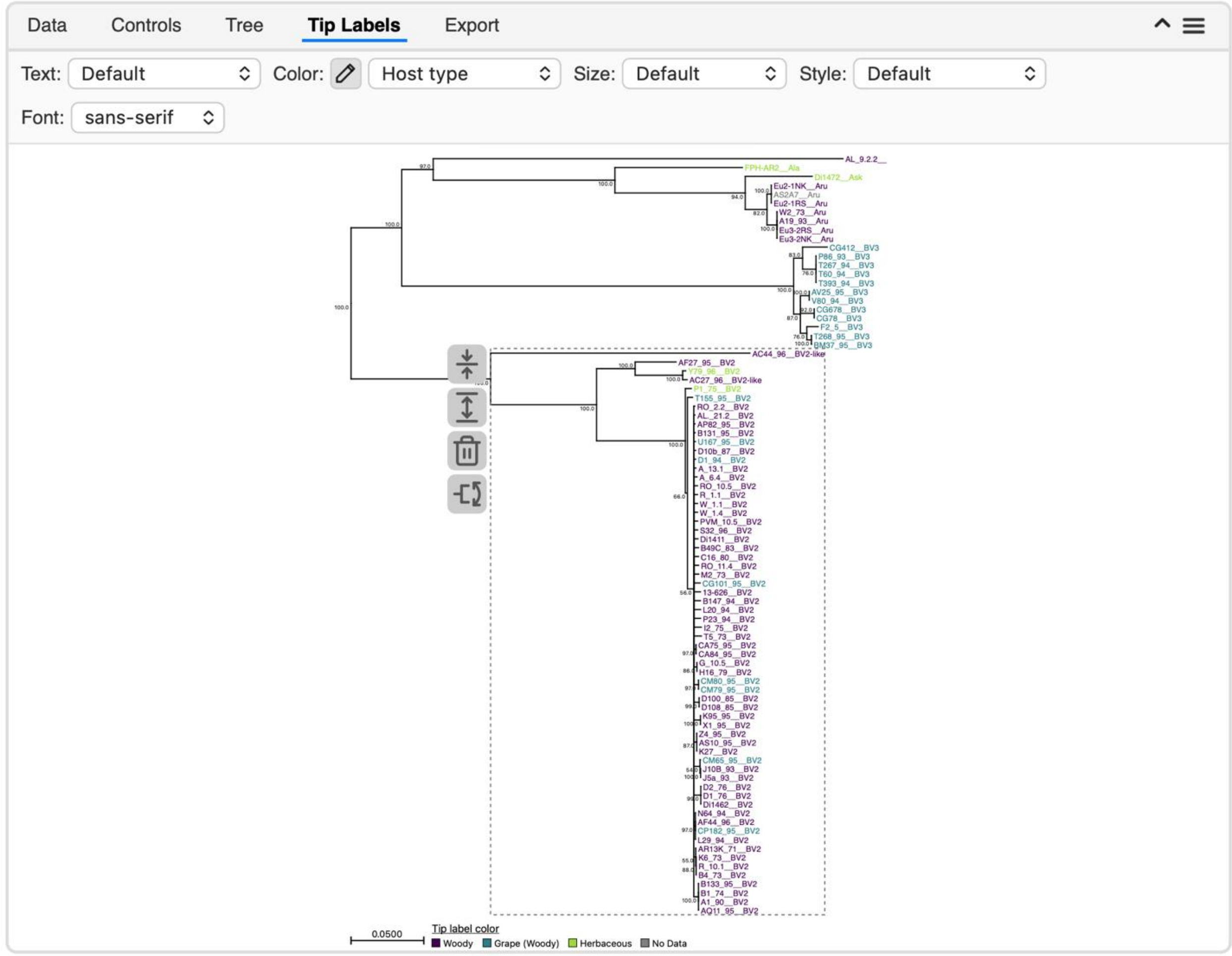

**Figure 2. Example `heat-tree` implementation in R for interactive, phylogenetic tree visualization.** The tree depicts currently recognized species within the *Agrobacterium-Rhizobium* complex based on the published MLSA maximum-likelihood phylogeny (Weisberg et al. 2020), with strain identifiers shown as tip labels and host type encoded by tip label color. A scale bar and legend indicating host-type categories are shown below the tree. The `heat-tree` interface is interactive and supports inspection and modification of phylogenetic trees through a collapsible control panel. An example visualization can be explored at https://grunwaldlab.github.io/heattree/articles/getting_started.html. Using this link, users may load other phylogenetic trees together with associated metadata (e.g., TSV/CSV files), adjust visual encodings based on metadata attributes, apply structural operations to selected branches (including flipping, removal, or subsetting), and view multiple trees concurrently within a session.

**Tables**

**Table 1: Comparison of** `heat-tree` with commonly used phylogenetic visualization tools (Letunic and Bork 2021; Rambaut 2009; Revell 2012; Yu et al. 2017).

| Feature | heat-tree | ggtree | iTOL | FigTree | Phylocanvas |
|---|---|---|---|---|---|
| Interactive | Yes | No | Yes | Partial | Yes |
| JavaScript API | Yes | No | Partial | No | Yes |
| R API | Yes | Yes | Partial | No | Yes |
| Python API | Yes | No | Partial | No | Partial |
| Web-compatible | Yes | No | Yes | No | Yes |
| Reproducible | Yes | Yes | Partial | No | Yes |
| Open source | Yes | Yes | No | Yes | Yes |
| Works offline | Yes | Yes | No | Yes | Yes |
| Metadata mapping | Yes | Yes | Yes | Partial | Partial |
| GUI | Yes | No | Yes | Yes | No |
| Embeddable | Yes | Yes | No | No | Yes |